# Massive Machine-type Communications in 5G: Physical and MAC-layer solutions

Carsten Bockelmann[*], Nuno Pratas[+], Hosein Nikopour[‡], Kelvin Au[♣], Tommy Svensson[†], Cedomir Stefanovic[+], Petar Popovski[+], Armin Dekorsy[*]

* {bockelmann, dekorsy}@ant.uni-bremen.de, University of Bremen, Department of Communications Engineering, Bremen, Germany

+ {nup, cs, petarp}@es.aau.dk, Department of Electronic Systems, Aalborg University, Aalborg, Denmark

♣ kelvin.au@huawei.com, Huawei Technologies Canada Co., LTD. Ottawa, Ontario, Canada

‡ hosein.nikopour@intel.com, Intel Labs, Wireless Communications Research, Santa Clara, CA, USA

† tommy.svensson@chalmers.se, Chalmers University of Technology, Gothenburg, Sweden

## 1 Abstract

Machine-type communications (MTC) are expected to play an essential role within future 5G systems. In the FP7 project METIS, MTC has been further classified into "massive Machine-Type Communication" (mMTC) and "ultra-reliable Machine-Type Communication" (uMTC). While mMTC is about wireless connectivity to tens of billions of machine-type terminals, uMTC is about availability, low latency, and high reliability. The main challenge in mMTC is scalable and efficient connectivity for a massive number of devices sending very short packets, which is not done adequately in cellular systems designed for human-type communications. Furthermore, mMTC solutions need to enable wide area coverage and deep indoor penetration while having low cost and being energy efficient. In this article, we introduce the physical (PHY) and medium access control (MAC) layer solutions developed within METIS to address this challenge.

## 2 Introduction

In the past years the development of a 5G vision led to the consensus that the latest generation of cellular communication systems will be driven by a number of newly emerging use cases [1], whereas the previous generations of cellular systems have been mainly designed towards increased spectral efficiencies to enable bandwidth-hungry applications for human users. Indeed, the FP7 Project METIS developed a 5G system concept in which this well-accepted chase for higher data rates is summarized by a service termed "Extreme Mobile Broadband" (xMBB) [2]. However, in addition to xMBB, new use cases and applications also necessitate fundamentally new services. The introduction of the new paradigm of machine-type communications (MTC), in addition to human-type traffic, poses significant challenges towards a unified radio solution.

The definition of machine-type communications is somewhat elusive, as it has to include a large variety of emerging concepts, such as the Internet of Things (IoT), Internet of Everything (IoE), Industry 4.0, Smart X, etc. Each adds new scenarios with differing assumptions and requirements, ranging from long-term environmental observation involving limited energy consumption, over smart cities with millions of sensors, to fully wireless factories with very strict requirements on latencies and reliabilities of the wireless connection. A 5G design needs to consider all this in order to really fulfill the role of a universal enabler for emerging and future industries.

The METIS project adopted the term "Machine-Type Communication" (MTC) from 3GPP to summarize all applications shaping the requirements on potential 5G solutions. Furthermore, MTC has been differentiated according to the two major challenges: (i) "massive Machine-Type Communication" (mMTC) and (ii) "ultra-reliable Machine-Type Communication" (uMTC) [2]. As the name suggests, mMTC is about massive access by a large number of devices, i.e., about providing wireless connectivity to tens of billions of often low-complexity low-power machine-type devices. Contrary to xMBB, where peak rates are prioritized, here the accent is on scalable connectivity for an increasing number of devices, wide area coverage and deep indoor penetration. A typical example of mMTC is the collection of the measurements from a massive number of sensors, such as smart metering. On the other hand, uMTC is about providing adequate wireless links for network services with rather stringent requirements on availability, latency and reliability. For the concept of uMTC, two important examples are Vehicle-to-X (V2X) communications and industrial control applications. Accordingly, the METIS MTC vision defined in [2] revolves around technologies that are able to solve these main challenges and jointly address further Key Performance Indicators (KPI), such as energy efficiency and cost to achieve a flexible and widely applicable cellular system.

A wide range of technologies on different layers of the ISO/OSI model are needed to provide satisfactory solutions for future MTC applications. In this article, we introduce the physical (PHY) and medium access control (MAC) layer technology solutions developed in METIS. While mMTC involves some challenges in downlink communication, e.g., efficient communication with actuator devices under duty cycles and sleeping patterns, the uplink represents a primary challenge due to the massive number of uncoordinated connections in mMTC and is the focus of the present article. We explain the need for new MAC and PHY technologies for massive and energy efficient access and present the PHY and MAC solutions developed in METIS.

## 3 Requirements and Design Challenges

The design on the PHY and MAC layer is fundamentally a question of trade-offs. Simply speaking, we trade-off performance (e.g., throughput, error rates, latency, etc.) vs. complexity and the required overhead (e.g., feedback or control signaling). In traditional cellular system design up to 4G, the focus has been on the support of high data rate downlink for human-type communication, using large packet sizes. The demand to support ever increasing data rates, while facing limited spectrum resources, has led to the development of very sophisticated PHY and MAC layer technologies that can better exploit the wireless channel (e.g., link adaptation with channel quality feedback, channel aware scheduling, adaptive beam-forming, etc.) and cope with transmission errors (e.g. hybrid ARQ, strong turbo codes, etc.). The large packets and high data rates in downlink play a crucial role: the control signaling is feasible only when it is negligible compared to the payload size; and the amount of feedback in the uplink is feasible only if it brings sufficiently high spectral efficiency gains for the downlink, without deteriorating the uplink traffic requirements.

In LTE, starting from the PHY over all the upper layers, large overhead is required to facilitate access, reliable transmission, authentication, security, etc. Furthermore, at the MAC layer, access and scheduling protocols demand reliable control information, which is often protected from errors using resource inefficient approaches, such as low modulation schemes and high redundancy coding. Consider the extreme example of an MTC device that wants to communicate a single byte of information. Already on the PHY layer, channel estimation requires pilots and link adaptation procedures require feedback information considerably exceeding the payload size. Therefore, a future 5G air interface for mMTC should target very lean control signaling approaches and PHY/MAC technologies with minimal overhead. To enable a lean signaling/overhead approach with constant or even reduced MTC transmitter complexity (similar to device complexity reduction efforts pursued in 3GPP for LTE-M and Narrow Band IoT [3]) an increase of the receiver complexity at the network side is expected, as shown later.

Changing any of the standard assumptions – from large to small packets, from high to low data rates or from a downlink focused to an uplink focused communication – results in a fundamental change of the design problem. Hence, mMTC focusing on the uplink communication of a massive number of low-rate devices, requires a completely different set of technologies than the ones designed to serve human-centric communications, such as the one present in LTE and previous cellular generations. To narrow the design space of potential mMTC technologies we have to look at typical requirements. Standard literature assumptions for mMTC are [4]:

- small packets potentially going down to a few bytes;
- large number of users, e.g. up to 300.000 devices in a single cell;
- uplink-dominated transmissions;
- low user data rates, e.g. around 10kb/s per user;
- sporadic user activity, e.g., mixed traffic models with period and event driven traffic;
- low complexity and battery constrained (low energy) MTC devices.

Given these requirements we can now analyze potential design choices and outline guidelines for an mMTC 5G air interface. We focus on the following overarching questions associated to provision of access for a massive number of devices in the uplink:

- Should we choose orthogonal or non-orthogonal medium access?
- Is grant-free or grant-based access control better suited for mMTC?
- How to enable energy efficient MTC devices through PHY and MAC technologies?

First, orthogonal medium access tightly couples the number of available resources to the number supportable users, whereas non-orthogonal medium access enables a certain degree of resource overloading at the cost of algorithmic complexity at the receiver. The latter seems favorable to support a large number of uplink users in a resource-efficient manner due to relaxed complexity constraints at the base station compared to the MTC device.

Second, grant-based access control requires a good prediction of the uplink requests, as well as additional control signaling or message exchanges to facilitate the granting of resources. In MTC, traffic patterns are partly unpredictable and sporadic due to uncoordinated sleeping cycles, making grant-based scheduled access design difficult and potentially inefficient. On the other hand, grant-free access requires only very low control overhead, but often suffers from collisions and low efficiency. However, with an appropriate collision resolution mechanism, grant-free access can be made highly efficient, but again at the cost of increased base station complexity. Consequently, grant-free access control seems favorable for mMTC.

Finally, the energy efficiency of MTC devices is strongly affected by the amount of overhead seen in the exchange of messages required before the data payload is transmitted successfully. Simply speaking, less frequent and shorter transmissions preserve energy. Therefore, low signaling overhead MAC protocols are one enabler for energy efficient MTC devices. These type of protocols, coupled with efficient PHY approaches, can then enable devices with long battery lives.

Note that the small packet transmissions, associated with MTC, lead to other challenges, in addition to the discussed overhead: (i) demand for higher resource granularity and (ii) channel coding for short block lengths. Even with a non-orthogonal grant-free access protocol in place, the frame-based organization of resources, as employed in LTE, may be too limited in terms of resource granularity to adapt to very short packets. The redesign of the current frame structure, with the purpose to increase the resource granularity and assignment flexibility with very low overhead, is essential to support a massive number of devices and enable low latency communications. Finally, the channel codes that are currently used in cellular networks, are designed to approach the channel capacity for long packets. Unfortunately, these coding techniques are not suitable for short packets, due to significantly degraded performance. Thus, novel channel codes designed for short packet lengths are an essential requirement for MTC; however, the detailed treatment of this issue here is out of scope.

In the following, we first discuss physical layer approaches that (i) enable massive access, which are Sparse Code Multiple Access (SCMA) and Compressed Sensing based Multi-User Detection (CS-MUD) and (ii) foster energy efficiency and low cost, which is Continuous Phase Modulation (CPM). Then we introduce two novel MAC approaches for mMTC, which are based on SCMA and CS-MUD, respectively.

## 4 Physical Layer Solutions

### 4.1 Compressed Sensing based Multi-User Detection (CS-MUD)

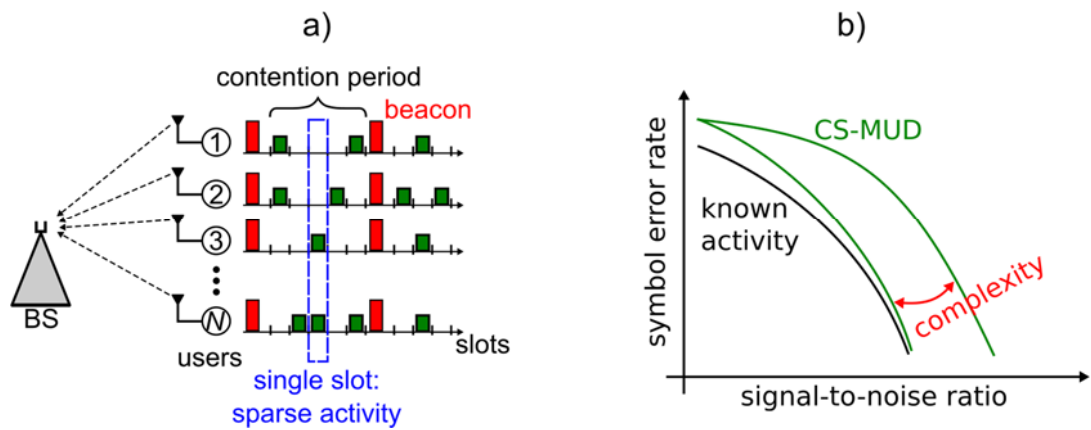

Figure 1 – (a) MTC uplink communication of N users to a base station (BS). A slotted ALOHA-based access protocol is assumed. Activity of multiple users in one slot leads to interference in terms of the physical layer. From the MAC perspective usually an unresolvable collision is assumed in such a case. (b) Qualitative error rate performance of CS-MUD compared to traditional detection with known activity.
4

As discussed in Section 3, random access protocols are strongly affected by collisions, i.e., multiple users access a resource concurrently and thus cannot be successfully detected and decoded. The outcome of collisions, though, is strongly dependent on the specific physical layer technologies: advanced receivers using interference cancellation can resolve a certain amount of collisions, whereas a combination with non-orthogonal medium access methods, like CDMA or SCMA (cf. section 4.2) boosts exploitation of collisions further. Consequently, Compressed Sensing based Multi-User Detection (CS-MUD) targets enhanced resource efficiency and number of served users through non-orthogonal random access in combination with a joint detection of user activity and data.

Figure 1a illustrates an uplink scenario with *N* users and a single base station. We assume a slotted ALOHA-based access scheme such that users are randomly active in the resulting contention period. Due to the random activity in each slot, only a subset of users is actively sending data. From a PHY perspective, this leads to estimation problems with sparsity, as the subset of active users is unknown and has to be estimated jointly with the user data. Compressive sensing (CS) focuses on similar estimation problems with sparsity constraints, which are widely applied in imaging and radar and have been only recently considered in the context of communications. Most importantly, CS provides a sound theoretical basis for novel advanced multi-user detection algorithms exploiting sparsity. Nonetheless, known CS algorithms require refinement to incorporate PHY layer assumptions, such as finite modulation alphabets, channel coding and error control of activity detection. Especially the latter is of high importance for activity estimation that is reliable in terms of missed detections and false alarms in order to suit the needs of MAC processing [5]. Solutions to the resulting requirements have been intensively studied within METIS, leading to a novel class of detection algorithms named "Compressive Sensing based Multi-User Detection" (CS-MUD) [6, 7]. These algorithms enable resource efficient, highly reliable random access in combination with non-orthogonal medium access schemes.

Figure 1b depicts a qualitative result of CS-MUD compared to an estimation problem with known user activity, which reverts to the standard data detection task. The variable gap between CS-MUD and known activity results shows the flexibility of the scheme. Depending on algorithmic complexity, different levels of side information such as frame structure, finite modulation alphabets or channel coding can be exploited to vastly improve the performance of CS-MUD. Clearly, the price for a low signaling solution with nearly the same performance as before is algorithmic complexity at the base station. On the one hand, careful design can decrease this complexity (cf. SCMA); on the other hand, exploiting more side information usually leads to more complex algorithms. The best trade-off will depend on the actual system and the specifics of the MAC layer processing.

To summarize, CS-MUD provides advanced collision resolution to serve a large number of users through random access with low overhead, but with performance comparable to grant-based schemes. CS-MUD requires a higher algorithmic complexity at the base station, compared to simple receiver often employed in scheduled systems, and is based on the assumption of sporadic traffic patterns and non-orthogonal medium access.

## 4.2 Sparse Code Multiple Access (SCMA)

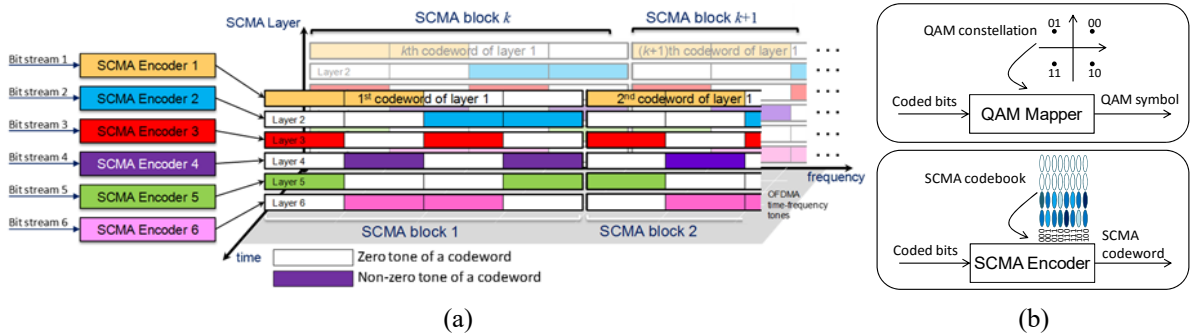

(a) (b)

Figure 2 – (a) 6 SCMA multiplexed layers carried over 4 OFDMA tones with 150% overloading. Codewords of 6 layers are selected from 6 layer-specific codebooks. Due to the sparsity, only 3 out of 6 layers collide over each OFDMA tone, (b) an SCMA encoder with a multi-dimensional codebook in comparison with a QAM mapper with a QAM constellation.

Sparse code multiple access (SCMA) [8] is a non-orthogonal multiple-access scheme in code domain that can support massive connectivity. As shown in Figure 2a, an SCMA encoder directly maps incoming bits of a data stream to SCMA codewords selected from a layer-specific codebook. Comparing the SCMA encoder to a QAM modulator as in Figure 2b, an SCMA encoder has the same functionality, but with multi-dimensional constellation points rather than traditional QAM constellations. Codewords of multiple SCMA layers are combined either at a transmit point in downlink or over the air and are carried over orthogonal channel resources such as Orthogonal Frequency Division Multiple Access (OFDMA) tones.

The main advantage of SCMA is achieved by proper codebook design. Comparing SCMA to traditional multi-carrier CDMA (MC-CDMA), we can see two differences: (i) standard MC-CDMA is equivalent to a repetition coding of QAM symbols; and (ii) QAM constellations exhibit a gap to the Shannon capacity. In SCMA, these two points are addressed through multi-dimensional codeword design [9]. A multi-dimensional lattice provides further degrees of freedom to better utilize the space over multiple tones and improve the distance spectrum of SCMA constellations. Thus, we achieve a coding as well as a shaping gain over CDMA repetition and OFDMA QAM constellations, respectively. It is important to note that Non-Orthogonal Multiple Access (NOMA), although having similar features as SCMA, it is unable to reach the same level of user overload and it is not suitable for massive uplink MTC scenarios.

Furthermore, the complex multi-dimensional codewords of SCMA allow a non-orthogonal design and hence overloading. Overloading increases the number of available links, which is beneficial for massive connectivity, but optimal multi-layer detection is challenging for practical systems. However, as opposed to non-orthogonal CDMA, near-optimal ML-like detection can be made practically feasible through proper design of the SCMA codewords and iterative message passing algorithm (MPA) [8].

Therefore, SCMA codewords are designed to be sparse [9]. Sparsity of codewords along with the low projection property helps the MPA receiver to exponentially reduce the detection complexity. Referring to Figure 2a, the system is 150% overloaded with 6 non-orthogonal layers, but due to the sparsity of the codewords only 3 of them collide over each OFDMA tone. As an example, each layer carries codewords of a layer-specific 16-point codebook. If the codebooks are not sparse (which is equivalent to CDMA with 16-QAM constellation), the total number of possible combinations over each tone is $16^6$, but thanks to the sparsity this can be reduced to $16^3$ as only 3 out of 6 codewords collide over each tone. As reported in [9] the effective size of the codebooks over each non-zero tone can be reduced from 16 to 9. It consequently reduces the complexity of MPA detection to $9^3$. Therefore, by combining sparsity and low projection techniques the original complexity of detection is extensively reduced by a factor of $\frac{16^6}{9^3} \sim 23014$ for this particular scenario [9].

In summary, the sparsity of codebooks, low projection techniques and blind detection with SCMA receiver make SCMA an attractive solution for uplink contention-based grant-free transmission in mMTC with a moderately higher complexity than a conventional linear receiver (Section 5.1).

## 4.3    Continuous Phase Modulation (CPM)

Many mMTC devices will be limited by their battery due to cost/space constraints, and should preferably use low cost amplifiers; therefore, energy efficiency of mMTC terminals and good coverage are more important than spectral efficiency. Constant envelope signals offer the possibility to use a non-linear cost-effective and power-efficient High-Power Amplifier (HPA) at the transmitter, because the HPA can be operated close to saturation without adding distortion. For this reason, constant envelope coded-modulation systems have traditionally been extensively used in, e.g., satellite links, early wireless standards (GSM), Bluetooth, and low rate long distance microwave radio links for cellular backhauling.

To illustrate the potential HPA efficiency gains, Figure 3 (left) shows how peak-to-average power ratio (PAPR) and the Raw Cubic Metric (RCM) (similar to 3GPP Cubic Metric) relate to the overall efficiency of a class E LDMOS High Power Amplifier (HPA) for a set of signals having different envelope properties [10]. As can be seen, the overall efficiency of the HPA for a CPM-like signal is substantially better than for Single-Carrier Frequency Division Multiple Access (SC-FDMA) signals used in the uplink of 3GPP LTE, e.g., 66% overall efficiency for CPM compared to 40-45% for SC-FDMA with a 0.1% clipping signal distortion constraint of the input samples (solid red curve). There are even larger gains of CPM compared to TDMA-OFDM (66% vs 35%).

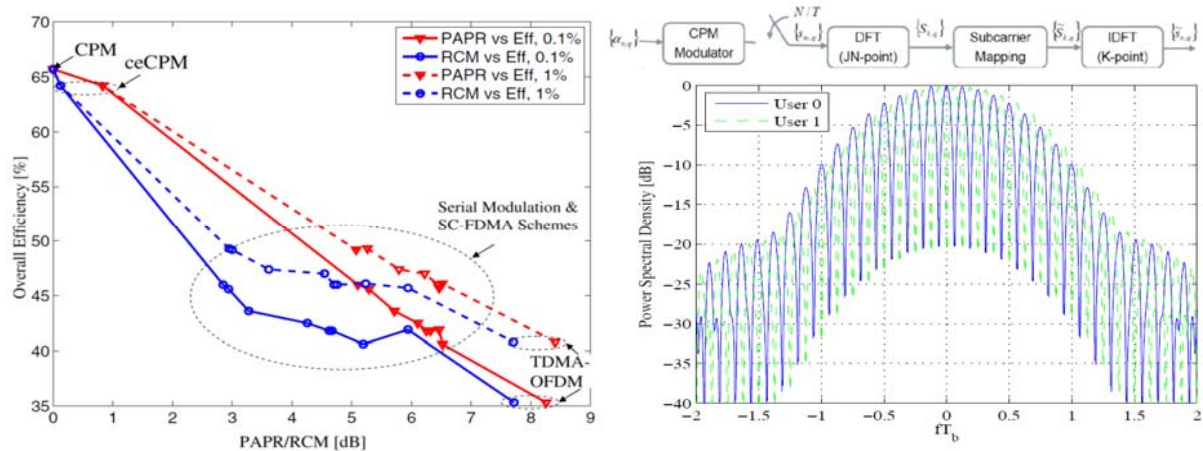

Figure 3 – Left: Maximum overall HPA efficiency versus mean PAPR and RCM of various modulated signals with 1% and 0.1% clipping level [10]. Right: Block diagram and power spectral density of an illustrative CPM-SC-FDMA scheme [12].

However, there is a trade-off. The lower the envelope variations, the more compact the signal space, leading to less sensitive receiver for a given receiver SNR. Thus, *constrained envelope* CPM (ceCPM) [11] has been proposed and developed based on a linear description of CPM. This approach allows the receiver sensitivity to be monotonically increased with the allowed envelope variation energy, using the same receiver architecture as in CPM. For example, it is possible to find a ceCPM scheme with an SNR gain of 2.5 dB in an AWGN channel over CPM at symbol error probability $10^{-3}$, under the same requirement on spectral confinement and spectral efficiency as for the best CPM scheme, with a PAPR penalty of about 1.5 dB. The ceCPM scheme is double as spectral efficient as Gaussian Minimum Shift Keying (GMSK) and can be detected with a low complexity Viterbi detector (using 8-16 states).

An alternative approach for frequency selective wideband multipath fading channels is illustrated in Figure 3 (right). This scheme uses sub-sampled CPM as precoder of SC-FDMA with Interleaved FDMA (I-FDMA) subcarrier mapping, treating the samples as generalized data symbols in SC-FDMA. Thus, a regular OFDMA transceiver is complemented with a precoder on the transmitter side, and a Viterbi decoder after soft symbol estimation on the receiver side. In this way, a constrained envelope signal can be conveyed using a spread spectrum approach over a frequency selective channel. Schemes obtained in [12] demonstrate that PAPR can be as low as a fraction of a dB. The users are frequency multiplexed in the I-FDMA fashion, thus harvesting frequency diversity towards fading and narrowband interference. In [12] it is shown that CPM-SC-FDMA can outperform convolutional encoded QPSK based SC-FDMA by up to 4 dB in end-to-end power efficiency, taking HPA power backoff into account (baseband power consumption not evaluated).

Thus, these promising constrained envelope coded modulation schemes exhibit useful properties for mMTC in both traditional narrowband single-carrier and wideband multi-carrier channelization scenarios, at the cost of more complex baseband for waveform processing.

## 5 Access Layer Solutions

### 5.1 Uplink SCMA Contention-Based Grant-free Transmission

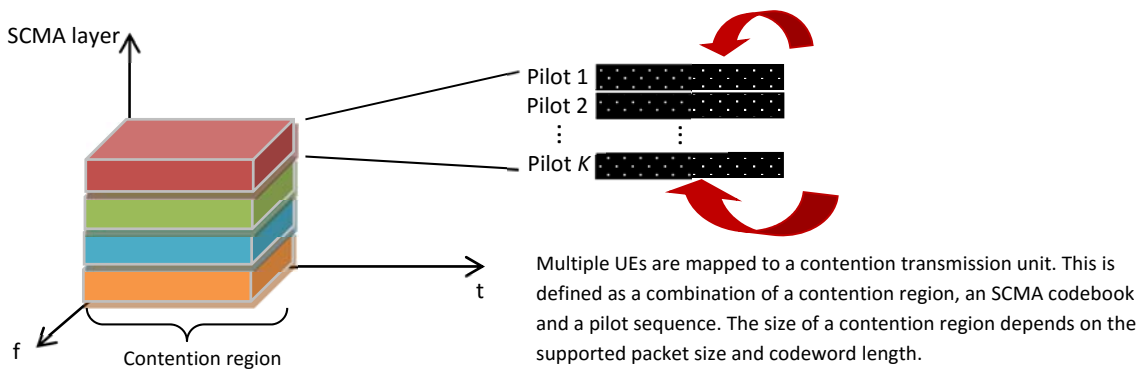

Multiple UEs are mapped to a contention transmission unit. This is defined as a combination of a contention region, an SCMA codebook and a pilot sequence. The size of a contention region depends on the supported packet size and codeword length.

Figure 4 – Uplink SCMA random access

As discussed in Section 3, random access is an attractive strategy to (i) remove or reduce the dynamic resource allocation overhead, and (ii) achieve latency reduction. A combination of random access with SCMA (cf. Section 4.2) is appealing in terms of overloading and scalability. The principle idea is illustrated in Figure 4. A contention transmission unit (CTU) is the basic radio resource defined as a combination of a contention region (time and frequency resource), an SCMA codebook and a pilot sequence. When a UE has data to transmit, it occupies the entire contention region to transmit codewords of one SCMA layer. Two different CTUs may share the same codebook but their pilot sequences are unique. Codebook reuse is allowed among UEs due to their statistically independent random channels on the uplink. Therefore, two users are colliding only if they have identical pilot sequences. A pilot collision prevents the receiver to estimate channels of colliding UEs for coherent MPA detection. To alleviate this, SCMA provides large number of links with high reliability, first due to overloading with sparse codebook design as discussed in Section 4.2, and secondly due to codebook reuse to generate a large pool of CTUs leading to low chance of collision. If the network still contains too many users, they need to contend for CTUs and collision can be resolved with conventional collision resolution procedures.

To adjust to dynamic changes, the number of contention regions and, hence, CTUs can be adapted based on the number of devices and collision statistics. Furthermore, SCMA offers scalability in terms of the number of codewords of an SCMA codebook, number of codebooks, codeword length, and the sparsity pattern of codebooks. These can be configured semi-statically based on the number of supported users, traffic volume of users, complexity of detection, coverage, reliability of links, and system outage. For example, longer codewords of length 8 with larger number of non-zero elements can provide better coding gain and hence better coverage, whereas sparser codewords (e.g. of length 4 with 1 or 2 non-zero elements) can tolerate further overloading to enable massive connectivity with feasible complexity of detection [13].

A crucial component of uplink SCMA grant-free transmission is blind detection. A regular MPA receiver, as discussed in Section 4.2, requires prior knowledge regarding the number of layers and their corresponding codebooks. However, in mMTC scenarios with random access, it is up to the SCMA receiver to recognize active layers and their data. This requires a modification of the original MPA receiver to provide joint activity and data detection just like CS-MUD does (cf. Section 4.1). By adding an all-zero codeword to model an inactive layer, a modified MPA receiver can detect inactive and active layers. In this case the activity and data of a layer are jointly detected by MPA [14].

To summarize, system-level simulations showed that up to 3 times more devices compared to contention-based OFDMA system can be supported for delay sensitive small packets at the cost of higher base station processing [13]. Moreover, SCMA can benefit from advances in CS-MUD (see Section 4.1) to further enhance the system's ability to tolerate more collisions.

## 5.2 Coded Random Access and CS-MUD

As previously discussed, the performance of random access protocols, such as slotted ALOHA (SA), is limited by the occurrence of collisions. In Coded Random Access (CRA) the theory and the tools of erasure-correcting codes are applied to enhance SA, drawing from the analogies between Successive Interference Cancellation (SIC) and iterative belief-propagation erasure-decoding. We consider a variant of CRA, denoted as Frameless ALOHA [15], in which: (1) the users contend on a slot basis, using predefined slot access probabilities, and where (2) the length of the contention period (in number of slots) is not a priori fixed, but determined on-the-fly so to maximize the throughput. This solution is focused on the efficient support of mMTC, being optimized for efficient resource use and to a high volume of served users.

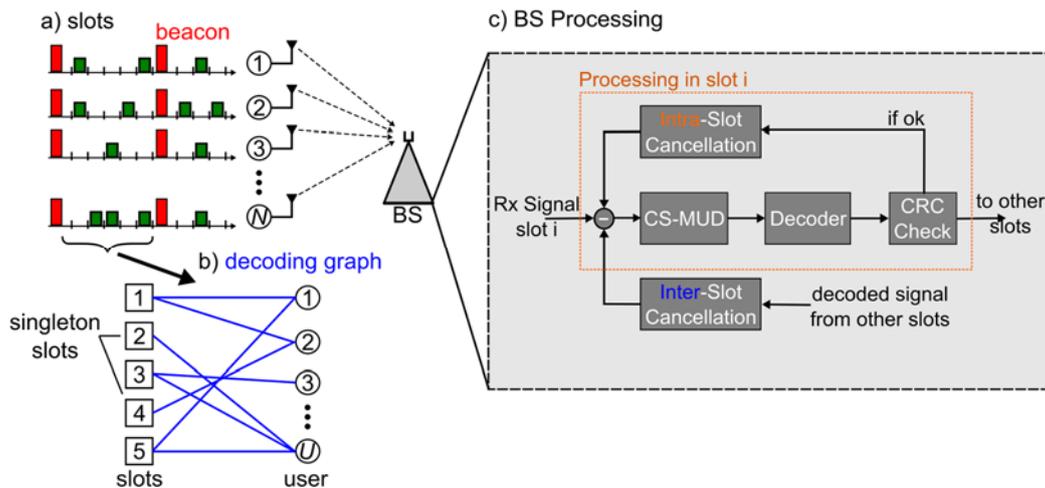

Figure 5 – Coded Random Access overview: a) frameless structure delimited by beacons; b) iterative belief propagation with execution of SIC on a graph at the inter-slot level; and c) intra-slot and inter-slot processing done at the base station.

One important CRA enabler is a physical layer capable of performing Multi-User Detection (MUD) and SIC, such as CS-MUD presented in Section 4.1. The combination of CRA and CS-MUD uses a unique data processing flow to successfully decode users over all received slots. The scheme starts with the transmission of a beacon, as depicted in Figure 5a, which initiates a MAC contention period. The UEs send replicas of the same packet in randomly selected slots, as shown in Figure 5b, using the slot access probability provided by the beacon. The contention period ends once the BS sends a new beacon. In each slot, as depicted in Figure 5c, the data processing takes place as follows: First, the PHY processing is facilitated by CS-MUD and standard FEC decoding to recover the UE packet, where a Cyclic Redundancy Check (CRC) check ensures the correct packet reception. If a UE's packet is correctly decoded, it will immediately be subtracted from the received signal at the current slot and the PHY processing may be repeated until no new UEs are successfully recovered. All successfully decoded UE packets are then stored for inter-slot interference cancellation, i.e., cancelling of replicas, which takes place once processing in the next slot starts and is performed based on belief-propagation on a graph, as shown in Figure 5b. In this way PHY and MAC processing work hand in hand to recover as many UE packets as possible, maximizing the throughput. This joint approach shows promising results in comparison to the LTE Rel. 11 baseline. Up to 10x the number of machine type devices can be supported for short packets, as shown in [7].

To summarize, the combination of CS-MUD and CRA provides an efficient PHY and MAC solution to solve sporadic mMTC allowing for asynchronous, sporadic mMTC uplink communications, where the active users, delays and channels are generally unknown. However, to enable the processing depicted in Figure 5c, the algorithmic complexity and the need for data buffering at the base station is significantly increased.

# 6 Conclusions

The design of 5G faces many challenges to provide the services of the future. We have presented a selection of candidate technologies developed in the FP7 project METIS for massive machine-type communication (mMTC). The main motivation of all presented solutions is the need for fresh approaches to solve the massive access problem in energy efficient and low cost ways. Of course, these are just a first step towards a complete solution and require further support in many ways. For example, the overall air interface design should allow for dynamic adaptation to different mMTC cell loads, while being well integrated with other services like xMBB. Currently, flexible waveform design enabling in-band mMTC channels seems to be a promising candidate in that respect. Also, higher layer considerations play a very important role to ensure lean signaling, e.g., by introduction of connectionless, one-shot transmission modes to enable longer sleep cycles, or traffic prediction approaches to efficiently support quasi-periodic traffic.

# 7 Acknowledgement

Part of this work has been performed in the framework of the FP7 project ICT- 317669 METIS, which is partly funded by the European Union. The authors would like to acknowledge the contributions of their colleagues in METIS, although the views expressed are those of the authors and do not necessarily represent the project.

The work of Č. Stefanović was supported by the Danish Council for Independent Research, grant no. DFF-4005-00281.

## 9  Biographies

Dr.-Ing. Carsten Bockelmannn received his Dipl.-Ing. (M.Sc.) degree in 2006 and his PhD degree 2012 both in electrical engineering and from the University of Bremen, Germany. Since 2012 he is working as a post doctoral researcher at the University of Bremen coordinating research activities regarding the application of compressive sensing/sampling to communication problems. His current research interests include compressive sensing and its application in communications contexts, as well as channel coding and transceiver design.

Nuno K. Pratas is an Assistant Professor on Wireless Communications at the Department of Electronic Systems, Aalborg University. He has been awarded twice the best student conference paper award and has been recognized as an exemplary reviewer for IEEE Transaction on Communications. His current research interests are on wireless communications, networks and development of analysis tools for Machine-to-Machine and Device-to-Device applications.

After two years of MIMO research at University of Waterloo, Ontario, Canada as a postdoctoral fellow, Dr. Hosein Nikopour joined Nortel Networks, Ontario, Canada in 2006. He was involved in WiMAX physical layer design as well as IEEE 16m and LTE standardization. In 2009, he joined Huawei Canada where he delivered wireless solutions for LTE and 5G cellular standards. Dr. Nikopour joined Intel Labs, Santa Clara, CA in July 2015 as a research scientist.

Kelvin Au is currently with Huawei Technologies Canada Co. Ltd. working on air interface research and standardization.  He received B.A.Sc. (1998) in Engineering Science and M.A.Sc. (2000) in Electrical Engineering from the University of Toronto, Canada. He joined Nortel Networks in 2000 focusing on PHY/MAC and radio resource management design for MIMO-OFDM systems. From 2008 to 2011, he was responsible for developing radio software in BlackBerry and was involved in several new product launches

TOMMY SVENSSON (S'98--M'03--SM'10) is Associate Professor in Communication Systems at Chalmers University of Technology in Gothenburg, Sweden, where he is leading the research on air interface and wireless backhaul networking technologies for future wireless systems. He received a Ph.D. in Information theory from Chalmers in 2003, and he has worked at Ericsson AB with access, microwave radio and core networks. He has co-authored three books and more than 120 journal and conference papers.

Čedomir Stefanović received the Dipl.-Ing., Mr.-Ing., and Ph.D.degrees in electrical engineering from the University of Novi Sad, Serbia. He is currently an associate professor at the Department of Electronic Systems, Aalborg University, Denmark. In 2014 he was awarded a individual postdoc grant by the Danish Council for Independent Research (Det Frie Forskningsråd). His research interests include coding theory, communication theory, and wireless communications.

Petar Popovski is a Professor at Aalborg University, Denmark. He received Dipl.-Ing. (1997)/Magister Ing. (2000) in communication engineering from Sts. Cyril and Methodius University, Skopje, Macedonia, and Ph. D. from Aalborg University (2004). He is a Fellow of IEEE. He currently serves as an Area Editor in IEEE Transactions on Wireless Communications and a Steering Committee member for IEEE Internet of Things Journal. His research interests are in wireless communications/networks and communication theory.

Prof. Armin Dekorsy is the head of the Department of Communications Engineering, University of Bremen. He received Dipl.-Ing. (FH) (1992) from Fachhochschule Konstanz, Germany, Dipl.-Ing. (1996) from University of Paderborn, Germany and Ph.D. (2000) from University of Bremen, Germany, all in communication engineering. He spent more than 10 years in industry at Deutsche Telekom AG, Bell Labs Europe (Lucent Technologies) and Qualcomm in leading research positions. He serves as member for ETSI, IEEE, VDE/ITG, and represents the University of Bremen at NetWorld2020 ETP.